\documentclass[%
pre,showkeys,
nofootinbib,  
amsmath,amssymb,
aps]{revtex4-2}

\usepackage{graphicx}
\usepackage{dcolumn}
\usepackage{bm}
\usepackage{siunitx}
\usepackage{xcolor}
\usepackage{balance}
\usepackage{siunitx}
\usepackage{bbding}
\usepackage[version=3]{mhchem}
\usepackage{mathptmx,amsmath,amssymb}
\usepackage[caption=false]{subfig}

\begin{document}
	\title{Weak non-linearities of amorphous polymer under creep in the vicinity of the glass transition}



		\begin{abstract}
		{Martin Roman-Faure, H\'el\`ene Montes, François Lequeux, Antoine Chateauminois \\
			{\it \noindent Soft Matter Science and Engineering (SIMM), CNRS UMR 7615, ESPCI Paris, PSL University, Sorbonne Universit\'e, F-75005 Paris, France}}
	
	The creep behavior of an amorphous poly(etherimide) (PEI) polymer is investigated in the vicinity of its glass transition in a weakly non linear regime where the acceleration of the creep response is driven by local configurational rearrangements. From the time shifts of the creep compliance curves under stresses from 1 to 15~\si{\mega\pascal} and in the temperature range between $T_g -10K$ and $T_g$, where $T_g$ is the glass transition, we determine a macroscopic acceleration factor. The macroscopic acceleration is shown to vary as $e^{-(\Sigma/Y)^n} $ with $n=2 \pm 0.2$, where $\Sigma$ is the macroscopic stress and $Y$ is a decreasing function of compliance. Because at the beginning of creep, the stress is homogeneous, the macroscopic acceleration is thus similar to the local one, in agreement with the recent theory of Long \textit{et al.} (\textit{Phys. Rev. Mat.} (2018) \textbf{2}, 105601 ) which predicts $n=2$. For larger compliances, the decrease of the of $Y$ is interpreted as a signature of the development of stress disorder during creep.

\end{abstract}

\maketitle


\section{Introduction}
Amorphous polymer and elastomer materials have linear mechanical properties which vary drastically around their glass transition~\cite{ferry_viscoelastic_1980}. At low temperatures, they exhibit an elastic modulus of the order of 1 GPa as a result of Van der Waal interactions between monomers~\cite{long_heterogeneous_2001}.
At high temperatures, their elastic modulus drops by three orders of magnitude towards about 1 MPa. In this domain, the elastic modulus is conferred by the entropy of the chains strands between cross-linking points for elastomers~\cite{rubinstein_elasticity_2002,rubinstein_polymer_2014}, or a so-called plateau modulus between entanglements for non-cross-linked polymers~\cite{ferry_viscoelastic_1980}. Similarly, mechanical properties at high strain also vary significantly with temperature. At high temperature, in the rubbery state, the macroscopic mechanical response still arise from chain entropy. As a consequence the macroscopic mechanical response remains approximately linear over a strain amplitude of 100\%~\cite{rubinstein_elasticity_2002}. On the other hand, in the glassy state, the mechanical response originates from molecular and Van der Waals interactions between close packed monomers. Then, non-linearity finds its origin in local rearrangements~\cite{argon_theory_1973,langer_shear-transformation-zone_2008,dequidt_heterogeneous_2016}. As a result, a macroscopic deformation of a few percent is enough to induce monomer rearrangements and to make the macroscopic response non-linear. This strong non-linearity is characterized by a stress overshoot known as the yield stress. However, in this work, we will focus on the crossover regime between glass and rubber behavior, an under-explored regime between $T_g-10K$ and $T_g$, $T_g$ being the glass transition temperature. In this regime, non-linearities do not appear as a yield stress, but instead as a tiny acceleration of the dynamics of rearrangements. Polymer glasses are highly disordered systems made up of nanometric domains, each of them with a specific relaxation time under thermal activation. These times are distributed over 4 orders of magnitude. This peculiar behavior is known as dynamical heterogeneities~\cite{ediger_spatially_2000}. Consequently, through the glass transition, the measurement time sweeps the distribution of local relaxation times which allows to reveal the respective role of domains with various relaxation times. In this study, we remain in a narrow temperature range, between $T_g-10$~\si{kelvin} and $T_g$ in which the experimental windows overlap the relaxation time distribution of the domains. As a consequence, domains relaxation can be either faster or slower than the measurement time, depending on their relaxation time. \\
Indeed, at the crossover between glassy and rubber states, domains rearrange under the effect of thermal agitation within a characteristic time $\tau$. However, under the action of a stress $\sigma$, the time required for rearrangement is accelerated~\cite{lee_deformation-induced_2009,lee_direct_2009,lee_molecular_2009,bennin_rejuvenation_2020,loo_chain_2000,perez-aparicio_dielectric_2016,perez-aparicio_dielectric_2016-1,kalfus_probing_2012,bending_measurement_2014,hebert_reversing_2017}, and becomes $f(\sigma)\tau$ where $f$ is a stress-dependent acceleration function, always smaller than unity,  which tends towards $1$ when $\sigma$ tends towards zero.  According to Eyring's classical model~\cite{eyring_viscosity_1936}, the acceleration function writes
\begin{equation}
	f=\exp^{-\sigma/Y},
	\label{eq:eyring}
\end{equation}
with $Y=kT/\textit{v}$, where $kT$ is the thermal energy and $\textit{v}$ is an activation volume. However, a more recent theory~\cite{dequidt_heterogeneous_2016,dequidt_mechanical_2012,long_dynamics_2018,conca_acceleration_2017} corroborated by local optical and mechanical measurements of segmental dynamics~\cite{lee_molecular_2009} and strain relaxation experiments~\cite{belguise_weak_2021} suggests that the acceleration function is rather
\begin{equation}
	f=\exp^{-(\sigma/Y)^2}
	\label{eq:long}
\end{equation}
with 
\begin{equation}
	Y=\sqrt{2 kT G'_0/\xi^3},
	\label{eq:xi}
\end{equation}
where $G'_0$ is the glassy elastic modulus and $\xi$ the size of the domains. Indeed, Eyring's expression~\cite{eyring_viscosity_1936} originates from a simple linear expansion with respect to stress of the value of the energy barrier  between two stable configurations of the monomers. In the Long~\textit{et al.} approach, the expression of the acceleration function is obtained using a different approach~\cite{long_dynamics_2018,dequidt_heterogeneous_2016,dequidt_mechanical_2012,conca_acceleration_2017} which is based on an estimate of the elastic energy required to cross the barrier.\\
Measuring and discussing the stress-dependence of this local acceleration function $f$ in amorphous polymers from macroscopic creep experiments in the vicinity of the glass transition is the main focus of our article. Obviously, a macroscopic acceleration function can be determined from a simple comparison of the experimental linear and non-linear mechanical response of the polymer. However, the relationship between this macroscopic acceleration function  and the local one  is difficult to establish because of the existence of stress heterogeneities. This is obvious in the case of internal stress as shown by Hasan and Boyce~\cite{hasan_energy_1993}, far below $T_g$, where huge amount of stress can be stored in sample with a zero macroscopic stress. However, even in the simple cases where the sample is annealed before the measurements, the development of stresses heterogeneities during under mechanical loading can make the comparison between local stress distribution and macroscopic stress very delicate. Indeed, we have shown recently that -in the same regime around $T_g$-, during the stress relaxation resulting from a step-strain, the macroscopic acceleration function can be very different from the local one~\cite{belguise_weak_2021} and more precisely that it can exhibit a different scaling. This is due to the fact that during a step-strain, the fast decay of the macroscopic stress during the measurement near $T_g$ makes the estimate of the local acceleration challenging. As discussed by Belguise~\textit{et al.}~\cite{belguise_weak_2021}, it requires the analysis of experimental data in the light of numerical simulations accounting for material disorder.\\
During creep, the changes in the local stress distribution within the polymer is somehow different from step-strain because the macroscopic applied stress remains constant. As a consequence, the average local stress remains constant. So, we expect that the local and macroscopic acceleration functions remain roughly similar as long as the stress disorder is weak. In this paper, we study the apparition of non-linearities during creep experiments using an amorphous poly(etherimide) (PEI) polymer as a model system. We limit ourselves to the close vicinity of the glass transition, between $T_g-10$~\si{kelvin} and $T_g$, and to situations where the macroscopic field of deformation has an amplitude typically smaller than 15\%, and where the stress is is smaller than $15$~\si{\mega\pascal}, \textit{i.e.} less than two hundredth of the glassy elastic modulus. In this strain range, we show that geometric non-linearities are negligible as compared to the one resulting from the stress-induced acceleration of configurational changes. This regime where a non linear mechanical response is observed in the absence of significant geometrical non-linearities will subsequently be denoted as the weak-non linear regime.\\
The paper is organized as follows: we first present the experimental protocol and the creep measurements carried out both in the linear and in the weakly non linear regimes. Then, we deduce the macroscopic acceleration. We subsequently discuss the results and establish that the local acceleration function is $f=e^{-(\sigma/Y)^n}$ with $n=2 \pm 0.2$ at the beginning of the creep. For larger value of the elapsed time we observe that the exponent remains close to $2$, but that the prefactor $Y$ decreases and we relate this observation to the consequences of increasing stress disorder as creep proceeds.\\
%
%
\section{Experimental details}
\subsection{Poly(etherimide) polymer}
This study has been carried out using injection molded poly(etherimide) (PEI) polymer sheets (Ultem 1010, Sabic) which were supplied by Arkema. The glass transition temperature of the used amorphous PEI grade is $T_g=213$~\si{\celsius} as measured by DSC at 10~\si{\celsius\per\minute} and its number molar mass is $M_n$=2~$10^4$~\si{\gram\per\mol}. The molar mass between entanglement, $M_e=$3.4~$10^3$~\si{\gram\per\mole}, was determined in a previous study~\cite{belguise_phd_2020} from measurements of the rubbery shear modulus ($G_R=1.6 \pm 0.2$~\si{\mega\pascal}) of the investigated PEI polymer. As detailed in Supplementary Information SI1, a master curve giving the shear linear viscoelastic properties of the polymer was established at $T_{ref}=213$~$\si{\celsius}$ for the purpose of a comparison of the associated shift factors to that determined from the creep compliance data (see below) between $Tg$ and $T_g+15$~\si{\celsius}.\\
\subsection{Creep measurements} 
Creep experiments in tensile mode were carried out using a MCR702 rheometer (Anton Paar, Austria). Here, we focus on the creep behaviour in a weakly non-linear regime where the applied stress is maintained below 15~\si{\mega\pascal} and the strain below than $15~\%$). The experimental justification for this threshold is provided in the last part of the paper. It especially ensures that geometrical non linearities can be neglected.\\
In this regime, a relevant comparison between the linear and non linear creep responses requires that the compliance $J(t,\Sigma)=\epsilon(t)/\Sigma$ (where $\epsilon(t)$ is the measured strain and $\Sigma$ is the macroscopic applied stress) is determined within an accuracy less than a few percent. Such a requirement prevents any comparison based on experiments carried out using different samples. Indeed, their mechanical responses can vary by as much as $10\%$ as a result of small uncertainties in their shapes or of small misalignment's in the grips of the rheometer. In order to avoid such an experimental scatter, the following procedure was developed: (\textit{i}) the creep master curve of a given specimen is first determined in the linear regime from successive measurements at $\sigma=1$~$\si{\mega\pascal}$ and at increasing temperatures from $198$ to $213~\si{\celsius}$ by 3~\si{\celsius} steps (we have checked experimentally that a 1~$\si{\mega\pascal}$ stress lies in the linear regime); (\textit{ii}) the non linear response of the \textit{same} specimen is subsequently measured at a given stress level and at increasing temperatures from $198$ to $213~\si{\celsius}$ while \textit{it remains fixed within the grips of the rheometer}. This procedure is repeated with a different specimen for each of the considered non-linear stress levels in the range 3-15~$\si{\mega\pascal}$. As explained in the next section, the relevant results for the determination of the acceleration function are limited to the range $204$ to $213~\si{\celsius}$. Indeed, such a procedure requires that the thermo-mechanical history of the specimen in between two successive creep steps is erased. For that purpose, the specimen is annealed after each creep sequence under a zero applied stress at $T=T_g-3$~\si{\celsius}. As detailed in Supplementary Information SI2, the duration of the recovery step was fixed in order to ensure that the residual strain is negligible with respect to the deformation achieved during the subsequent creep sequence. After each isothermal strain recovery step, the specimen is subsequently allowed to cool down to the prescribed temperature of the next creep sequence in about 400~$\si{\second}$. Then, a 600~$\si{\second}$ isotherm is observed before the application of stress in order to achieve a uniform temperature within the sample.\\
Creep experiments were carried out using dog-bone specimens which were milled from 2~\si{\milli\meter} thick PEI plates. The center section of the dogbone was 1~\si{\milli\meter} wide and 10~\si{\milli\meter} long.
The elongation of the sample was measured with a \si{\micro\meter} accuracy under constant applied forces ranging from 2 to 30 \si{\newton}, \textit{i.e.} under a constant nominal stress (from 1 to 15 \si{\mega\pascal}). Any geometrical non linearity being neglected, the compliance is calculated using the nominal stress both in the linear and non linear regimes. The rheometer was equipped with a windows which allowed for optical observations of the specimen within the chamber using a zoom lens and a CCD camera. These observations (see Supplementary Information SI4) did not show any evidence of necking or crazing of the specimen within the resolution of the optics ($\approx 10$~\si{\micro\meter}).\\
%
%
\section{Experimental results}
\subsection{Creep compliance}
\begin{figure} [!ht]
	\centering
	\includegraphics[width=0.5\linewidth]{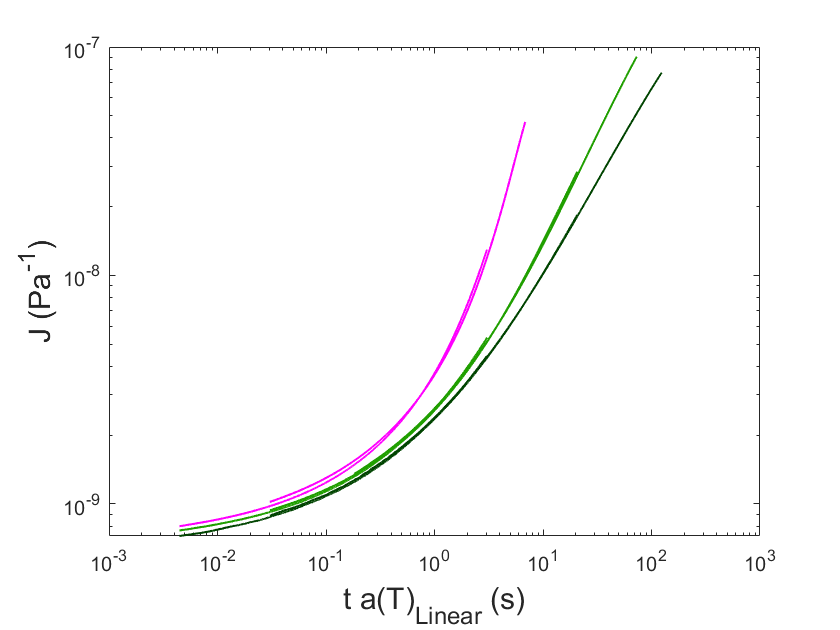}
	\caption{Creep compliance, $J_L$, in the linear regime (for $\Sigma = 1~\si{\mega\pascal}$, in black) and $J_{NL}$, in the nonlinear regime, for $\Sigma=5$~\si{\mega\pascal} (green) and $\Sigma=10$~\si{\mega\pascal} (magenta) as a function of the reduced time $a_Tt$ at the reference temperature $T_{ref}=213\si{\celsius}$. The master curves in the non linear regime were established using the shift factors $a_T$ determined in the linear regime. Only isotherms obtained at $T-T_g>-15\si{\celsius}$ are reported.}
	\label{fig:J_lin_nonlin}
\end{figure}

As a reference, the master curve giving the creep compliance $J_L$ in the linear regime (black line in Fig.~\ref{fig:J_lin_nonlin}) was established at $T_{ref}=T_g=213\si{\celsius}$ from the superposition of the creep compliance isotherms at $\Sigma=1 \si{\mega\pascal}$. As detailed in SI1, the corresponding shift factors $a_T$ are in perfect agreement with those deduced from the viscoelastic shear modulus measurements.\\
%
In addition to the linear compliance, the non-linear compliance $J_{NL}(\Sigma,a_Tt)$ is also reported in Fig.~\ref{fig:J_lin_nonlin} for $\Sigma=$5 and 10~\si{\mega\pascal}. These master curves were established using the shift factors $a_T$ which were determined in the linear regime. As shown in Supplementary Information SI3, the isotherms for $T-T_g<-15\si{\celsius}$ are affected by thermal history effects which are outside the scope of this study. They were thus discarded from Fig.~\ref{fig:J_lin_nonlin}.\\
The main effect of applied stress in the non linear regime is to induce an overall increase in the compliance values that we interpret, as detailed below, as a shift of the compliance to short times. The imperfect overlap of the isotherms indicates that the time-temperature superposition obeyed in the linear regime only poorly applies to the non-linear regime, which was noticed previously~\cite{oconnell_large_1997,oconnell_no_2002}. However, the amplitude of the resulting scatter is much smaller than the magnitude of the observed time shift with respect to the linear regime. As a consequence, we will neglect in what follows any effect of temperature in the description of the acceleration of the creep behavior under stress.\\
\subsection{Weak non-linear regime}
As detailed below, the relevant experiments for our study are restricted to a temperature between $T_g-10K$ and $T_g$. The stress is also limited to $15$~\si{\mega\pascal} in order to avoid any damage to the sample~\cite{djukic2020}. In addition, only strain values below $0.15$ are considered in the analysis of the acceleration of creep for the following reason. Any strain leads to a to a reduction in the cross-sectional area of the sample. During a tensile creep experiment, \textit{i.e.} under a constant applied force, the true stress sustained by the specimen is thus expected to increase as a result of this decrease in the sample cross-section. We have checked that this difference between the nominal and the true stress has no impact on our results for strain smaller than $0.15$ in the investigated temperature range. More generally, non-linearities in the mechanical response of materials are generally considered to result from a combination of both \textit{geometrical} and \textit{materials} non-linearities. Within the framework of finite strain theories, the former are described by the deformation gradient tensor which account for the non-linear relationship between strain and displacements, while materials non-linearities are embedded in the material's constitutive behavior law~\cite{dvorkin2006}. Here, keeping the creep strain below $0.15$ ensures that most of the geometrical non-linearities are avoided. We denote as the weak non-linear regime this peculiar set of strain and stress conditions, where the observed non-linearities originate only from the material's properties and not from any other source of non-linearities (geometry, occurrence of damage).

\subsection{Acceleration function}
\begin{figure} [!ht]
	\centering
	\includegraphics[width=0.5\linewidth]{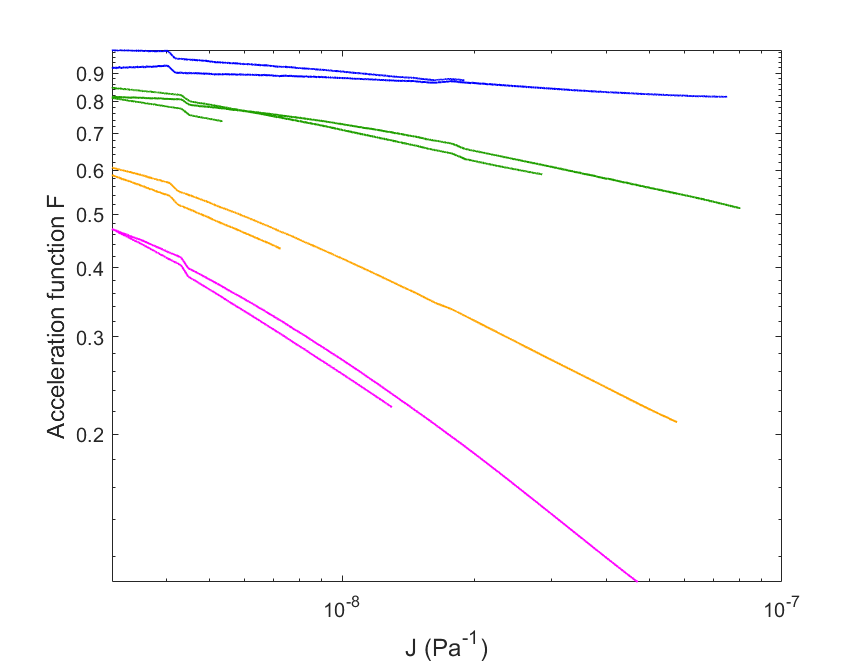}
	\caption{Macroscopic acceleration function $F$ as a function of the compliance $J$ for increasing values of the applied stress $\Sigma$. $\Sigma=$~3~\si{\mega\pascal} (blue), 5~\si{\mega\pascal} (green), 8~\si{\mega\pascal} (orange) and 10~\si{\mega\pascal} (magenta).  Data with $J \lesssim 3.10^{-9}$~\si{\per\pascal} where discarded from the figure as they are strongly influenced by the thermo-mechanical history of the specimen.}
	\label{fig:F_macro}
\end{figure}

\begin{figure}[htb]
	
	\subfloat[\label{subfig:3a}]{%
		\includegraphics[width=0.4\columnwidth]{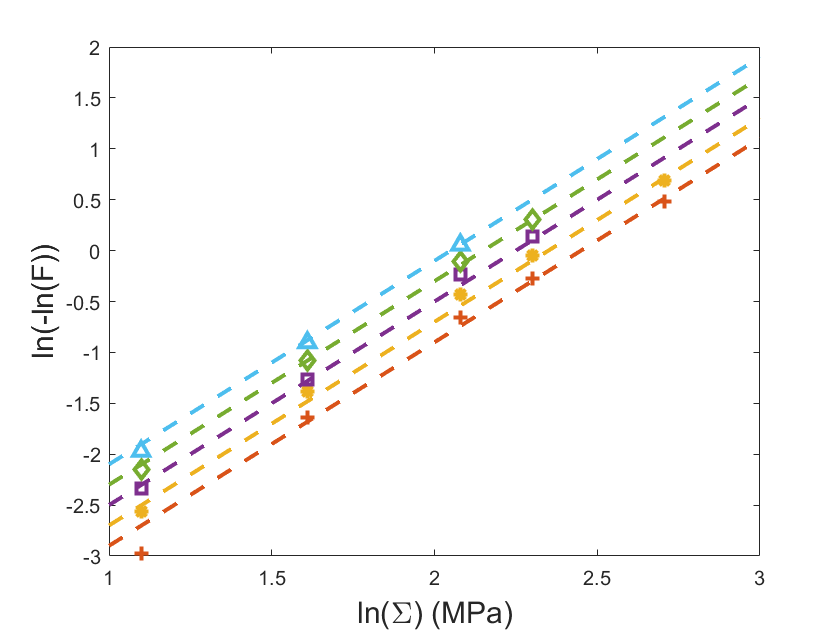}%
	}\hfill
	\subfloat[\label{subfig:3b}]{%
		\includegraphics[width=0.4\columnwidth]{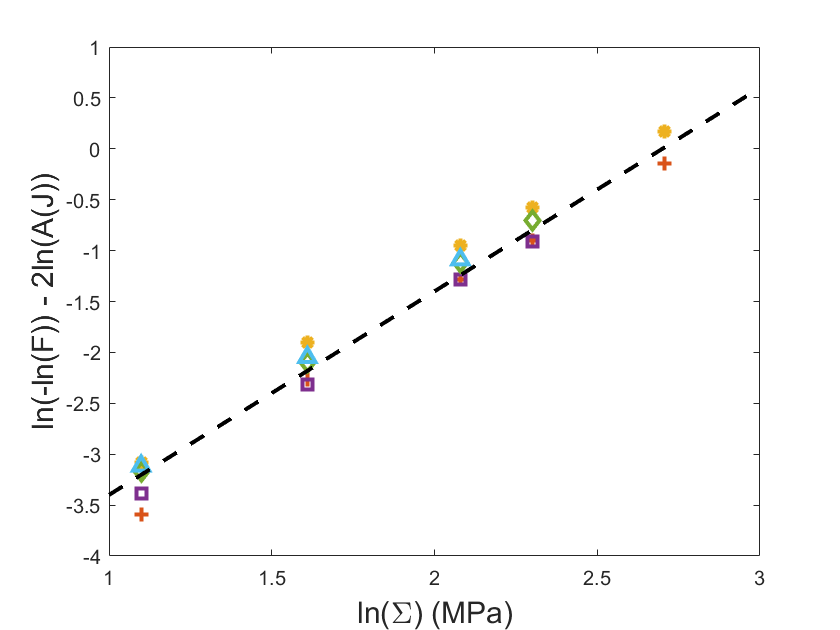}%
	}	
	\caption{Stress dependence of the macroscopic acceleration function $F$. (a) $\ln(-\ln(F))$ as a function of $\ln(\Sigma)$ for increasing values of the compliance $J$.  Dotted lines correspond to a linear fit of the data for every $J$ values while setting $n=2$ for the slope. (b) $\ln(-\ln(F)) - 2\ln(A)$ as a function of $\ln(\Sigma)$, where $\ln(A)=\ln(\frac{\Sigma}{Y(J/J_0)})$. From bottom to top: (\textcolor{red}{+}) $J=$~3.04~10$^{-9}$, (\textcolor{yellow}{\textbullet})~4.60~10$^{-9}$, (\textcolor{purple}{$\square$})~6.98~10$^{-9}$ (\textcolor{green}{$\diamond$})~1.06~10$^{-8}$ and $\triangle$)~1.61~10$^{-8}$~\si{\per\pascal}. The dotted line is a linear fit.}
	
\end{figure}

\begin{figure}[htb]
	
	\subfloat[\label{subfig:4a}]{%
		\includegraphics[width=0.4\columnwidth]{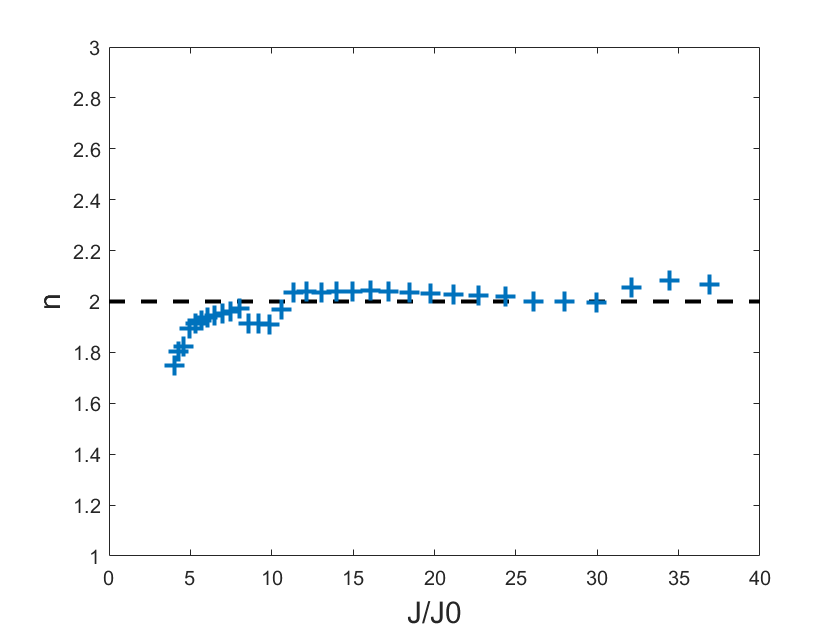}%
	}\hfill
	\subfloat[\label{subfig:4b}]{%
		\includegraphics[width=0.4\columnwidth]{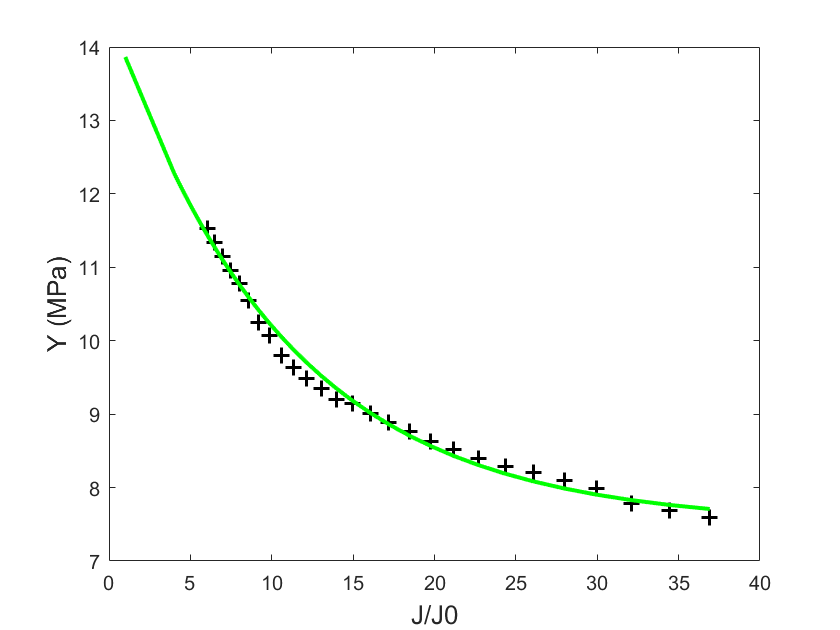}%
	}
	\caption{Values of the parameters of the fit of experimental $F(\Sigma,J)$ data to Eq.~\ref{eq:F_1}. (a) exponent $n$ as a function of normalized compliance $J/J_0$. The dotted line corresponds to the theoretical value, $n=2$. (b) $Y$ as a function of $J/J_0$. The green line corresponds to a fit of experimental data to an empirical decreasing exponential function, $Y=7e^{-\frac{J}{10.5 J_0}} + 7.5$.} 
\end{figure}

We now turn to a quantitative description of the time shift of the creep response in the non linear regime from the definition of an acceleration function. In what follows, we choose to define a macroscopic acceleration for creep $F$ as
\begin{eqnarray}
	J_{NL}(\Sigma,t)= J_L\left(\frac{t}{F}\right) .
\end{eqnarray} 
where $\Sigma$ is the applied macroscopic stress. Obviously, as all the measured creep compliance functions are monotonic and varying in the same range, $F$ can always be determined. The  choice of the definition of $F$ will be justified in the next section.\\ 
The acceleration is thus defined as the coefficient by which each point of the $J_{NL}(\Sigma,t)$  curves has to be shifted horizontally to superimpose to the point with the same compliance on the linear $J_{L}(t)$ curve. As it can be seen in Fig.~\ref{fig:J_lin_nonlin}, the macroscopic acceleration function $F$ is a not only a function of $\Sigma$ but also of $J$ and should be written $F(\Sigma,J_L)$. This function $F$ is plotted in Fig.~\ref{fig:F_macro} as a function of the compliance $J_L$ for various values of the applied stress $\Sigma$. As detailed in Supplementary Information SI3, it turns out that the measurement of acceleration are complicated by non-linear physical ageing  when $J \lesssim 3.10^{-9}~\si{\per\pascal}$ which corresponds to the lowest temperatures  $T<T_g-10 K$. A description of these aging effects on non linear creep is beyond the scope of this study and we will therefore restrict in what follows to the description of the acceleration function for compliance greater than $3.10^{-9}~\si{\per\pascal}$ for which we showed that the effects of ageing on the compliance curve are negligible as compared to that of stresses in the non linear range (see SI3).\\
%
As shown in Fig.~\ref{fig:F_macro}, the acceleration function is decreasing with increasing compliance, an effect which is enhanced at large stresses. Within experimental accuracy, we were not able to detect any significant effect of the temperature on $F$ for the considered stress and temperature ranges.\\
Taking inspiration from the theoretical expression derived by Long~\textit{et al.}~\cite{long_dynamics_2018,dequidt_heterogeneous_2016} for the local acceleration function $f$, we checked whether the experimental macroscopic acceleration function $F$ can be fitted using the following expression
\begin{equation}
	F=e^{-\left(\frac{\Sigma}{Y(J_L)}\right)^n}.
	\label{eq:F_1}
\end{equation}
It appears that this function is able to describe properly our data. In Fig.~\ref{subfig:3a}, $\ln(-\ln(F))$ is plotted  as a function of $\ln(\Sigma)$ for given values of the compliance $J_L$. Remarkably, a linear plot is obtained for all the $J_L$ values with an exponent $n=2 \pm 0.2$ which is independent on $J_L$ (see Fig.~\ref{subfig:4a}).
%
%

%
%
Conversely, the vertical shift of the $\ln(-\ln(F))$ \textit{vs} $\ln(\Sigma)$ plots when $J$ increases indicates that the parameter $Y$ of the fit is a function of the compliance. This dependency is further considered in Fig.~\ref{subfig:4b}, where $Y$ values deduced from the fit to Eq.~\ref{eq:F_1} (with $n=2$) are plotted as a function of the normalized compliance $J/J_0$, where $J_0=J(t=0)$.
%
%
%
In what follows, we discuss the possible origin of the observed decrease in $Y$ at increasing $J/J_0$ from a consideration of the development of stress inhomogeneities at the scale of nanometric domains.\\
%
%
%
\section{Discussion}
\subsection{Origin of the mechanical non-linearities }
We now discuss the dependence of $F$ with $\Sigma$ and $J_L$.  For that purpose, we briefly recall the basis of the current description of flow in amorphous polymers.After annealing, before the application of the creep stress, the local stress is zero everywhere. Moreover, at the local scale and before the occurrence of any relaxation - or plastic deformation or rearrangement - the response of the material is elastic and linear. Indeed, the macroscopic yield strain of amorphous polymers far below $T_g$ is about a few percent, thus suggesting that the strain at which rearrangements are initiated is also about a few percent. At such low strain levels, the elastic response of the domains is expected to be linear. This implies that the non-linear response originates only from the rearrangement of the local domains, as hypothesized in most of the models of amorphous plasticity~\cite{langer_shear-transformation-zone_2008,argon_theory_1973,dequidt_heterogeneous_2016}. Noticeably, some work by Wang and coworkers mentioned that long-ranged interactions through the chain connectivity can play a significant contribution in the extensional mechanical ductility of polymer glasses in the yield regime~\cite{wang_2014,cheng_2014}. However, such effects are not likely to play a dominant role in the development of stress heterogeneities in the weakly non linear regime that we are investigating.\\
As a conclusion, we consider that the only source of non linearity is the modification of the relaxation kinetics under stress.
\subsection{Acceleration at the onset of creep: a validation of the Long et al behavior}
At the beginning of creep, because of annealing, the stress field can be assumed to be homogeneous and the macroscopic acceleration must therefore be equal to the local one. Our data show that the macroscopic acceleration for $t \rightarrow 0$ writes $F= \exp\left(-(\Sigma/Y_{0})^{2\pm 0.2}\right)$ where $Y_0=Y(J=J_0)$. Hence, this result proves that the model suggested by Long~\textit{et al.}~\cite{long_dynamics_2018} provides a good description of the acceleration. This is indeed the main result of this work which confirms for the first time the validity of the theory of Long~\textit{et al.} from a direct mechanical measurement. Using Eq.~\ref{eq:xi} with $Y_0=14$~\si{\mega\pascal} allows to deduce the size $\xi=5.2$~\si{\nano\meter} of the domains which is in agreement with values reported in the literature~\cite{chen_segmental_2009,chen_theory_2011,tracht_combined_1999,merabia_heterogeneous_2002}. However, the dependence of $Y$ on $J_L$ still remains to be discussed.\\
\subsection{Dependence of $Y$ on creep compliance}
As explained above, the macroscopic creep behavior originates from the increasing number of relaxed domains as a result of plastic rearrangements, according to classical models. During creep, the stress field thus becomes more and more heterogeneous and the local stresses sustained by unrelaxed domains increase. Let's assume that the all the relaxed domains do no longer sustain any significant stress, which is a reasonable assumption in the case of the investigated weakly non-linear creep regime for which chain extensibility is not involved. We denote $\phi$ the volume fraction of these relaxed domains. During the course of linear creep, as the average or macroscopic stress $\Sigma$ remains constant, the macroscopic strain increases at an enhanced rate because most of the stress is carried by a decreasing fraction $1-\phi$ of unrelaxed domains. $J_L$  is thus an increasing function of $\phi$ only. The case of non-linear creep is very similar; the only difference is that that the rate of relaxation is accelerated. As a consequence, the rate of growth of the fraction $\phi$ of relaxed elements is enhanced. The macroscopic stress $\Sigma$ being sustained mainly by a decreasing fraction of domains, the acceleration function $F$ must therefore be a decreasing function of $\phi$ and thus of $J_L$. \\
To sum up, if, at the beginning of the creep when $t\rightarrow 0$, the local acceleration is equal to the macroscopic one, it decreases with time - or with increasing values of $J_L$ -  because the applied stress progressively localises in the slowest domains.  In the next section we attempt do describe more in detail the effect of stress heterogeneities on the variation of $Y$ from a mean field approach.\\
\subsection{Approximate mean field model}
\begin{figure} [!ht]
	\centering
	\includegraphics[width=0.5\linewidth]{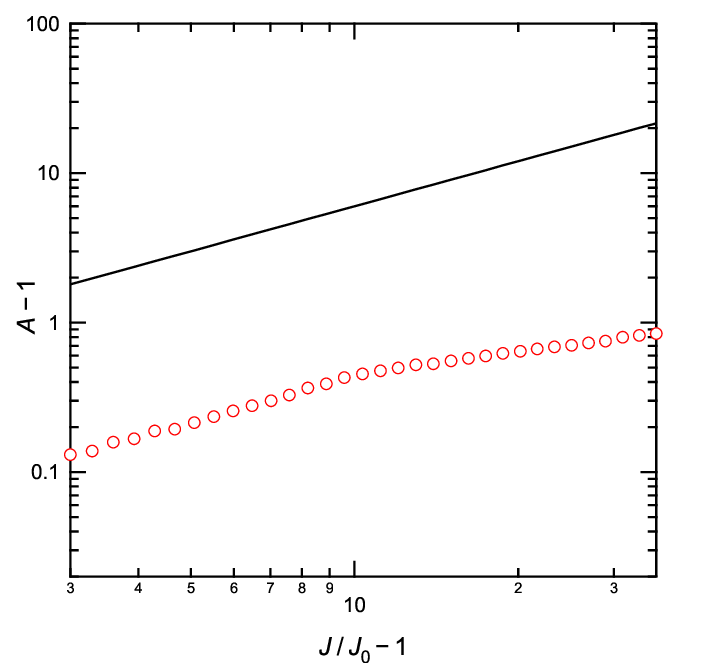}
	\caption{Stress amplification $A(J)-1$ as a function of $J/J_0-1$. The value of the amplification factor $A$ is determined using Eq.~\ref{eq:a_j} and $Y_0=14$~\si{\mega\pascal}. The black line corresponds to Eq.~\ref{eq:AJ}.} 
	\label{fig:AJ}
\end{figure}

\begin{figure} [!ht]
	\centering
	\includegraphics[width=0.6\linewidth]{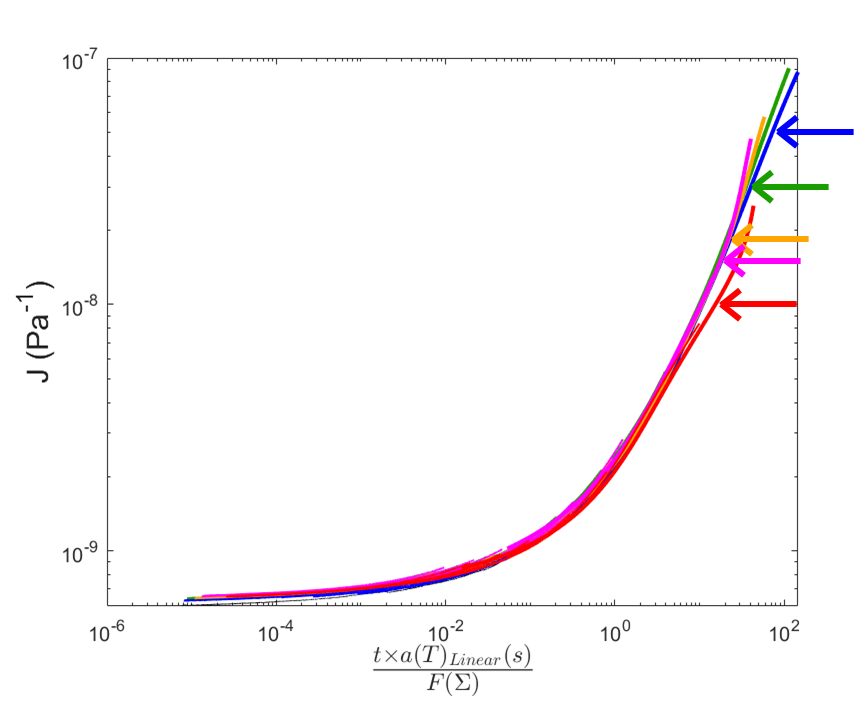}
	\caption{Master curve giving the compliance $J$ as function of the reduced time $a_T t \exp^{-\left(\frac{A(J)\Sigma}{Y_0}\right)^n}$. Applied stress $\Sigma$=3~\si{\mega\pascal} (blue), 5~\si{\mega\pascal} (green), 8~\si{\mega\pascal} (orange)  10~\si{\mega\pascal} (magenta) and 15~\si{\mega\pascal} (red). The arrows with the corresponding colors indicate the 15\% strain limit used when processing the data.} 
	\label{fig:Jmaster}
\end{figure}

The main difficulty that one may encounter in relating the macroscopic acceleration to the local one arises from the complex heterogeneity of the stress and strain field~\cite{masurel_role_2015}. It is out of the scope of this paper to solve completely the theoretical problem of the sequence of domains' relaxation and of the associated stress field redistribution. However, we develop in what follows a very crude description of the behavior of the polymer on the basis of a mean field approximation. We will assume that all the relaxed domains (with a volume fraction $\phi$) do no longer sustain any stress and behave like cavity while all non-relaxed domains (with a volume fraction $(1-\phi)$) bear the same local stress $\sigma$ which is amplified with respect of the applied stress $\Sigma$ by the same amplification factor. This approximation obviously leads to an overestimation of the stress carried by the unrelaxed domains.
In order to estimate this stress and the associated amplification factor, we use the Palierne's model~\cite{palierne_linear_1990} for the modulus of elastic systems with inclusions. This model assumes that the system is incompressible, which is a reasonable assumption in the investigated stress regime, as validated numerically in a previous study~\cite{belguise_weak_2021}. We also set all the interfacial stresses of the Palierne's model to zero. Let $G_0$ be the glassy modulus of the individual unrelaxed domains. For our system with cavities within a matrix of unrelaxed domains, the macroscopic modulus derived from Palierne's model writes (Eq.~4.2 in reference~\cite{palierne_linear_1990}) 
\begin{equation}
	G(t)=G_0 \frac{1-\phi}{1+2/3\phi}.
	\label{eq:G}
\end{equation}
The compliance can thus be expressed as
\begin{equation}
	J(t)=J_0\frac{1+2/3\phi(t)}{1-\phi(t)}.
	\label{eq:J_phi}
\end{equation}
Meanwhile, the local stress $\sigma$ is sustained by unrelaxed zones only. Within the framework of the assumption that all the non-relaxed domains are identical mechanically, we have therefore %
\begin{equation}
	\sigma (1-\phi)=\Sigma,
	\label{eq:stress}
\end{equation}
as the macroscopic stress is an average of the local one over the unrelaxed domains. Hence, the average stress amplification factor $A=\sigma/\Sigma$ for the remaining unrelaxed domains writes
\begin{equation}
	A=\frac{2}{5}+\frac{3}{5}\frac{J(t)}{J_0}
	\label{eq:AJ}
\end{equation}
According to this very crude approximation, the stress amplification factor $A$ should therefore increase linearly with the compliance at the beginning of the creep. In our approximation, for which the microscopic stress distribution in the unrelaxed domains is replaced by a single value $A(J)\Sigma$, we therefore expect to have
\begin{equation}
	F(\Sigma,J)\simeq f(A(J)\Sigma)
	\label{eq:F_2}
\end{equation}
According to Eq.~\ref{eq:F_1} and \ref{eq:F_2}, the stress amplification factor can thus be expressed as
\begin{equation}
	A(J)=Y_0/Y(J)
	\label{eq:a_j}
\end{equation}
where $Y_0=Y(J_0)$. As shown in Fig.~\ref{subfig:4b}, it appears that $Y(J/J_0)$ can be fitted reasonably with a decreasing exponential function from which we obtain $Y_0=14~\si{\mega\pascal}$. The stress amplification factor $A$ was thus determined using this value and Eq.~\ref{eq:a_j}. From Fig.~\ref{fig:AJ} where $A-1$ is reported as a function of $J/J_0-1$. Although it captures the trends of the experimental $A(J)$ relationship, it turns out that our mean field approach overestimates stress amplification by a factor of about ten. This overestimate originates likely in the strong assumption that the relaxed domains do not carry anymore stress, which is certainly not the case. The relaxed domains should behave like elastic domains but with a plastic deformation. Moreover, our mean field approach probably discards many specific features of stress localization which result from the mechanical coupling between domains. Indeed, Finite Element simulations of the mechanics of amorphous polymers using a more refined stochastic continuum mechanics description of the disorder show that stress amplification occurs within band oriented at 45~deg. with respect to the applied stress~\cite{masurel_role_2015}. The development of such bands evidences collective relaxation mechanisms which are not accounted for in our mean field approximation. Nevertheless, the later, even if it overestimates the stress localization by a factor of about ten, allows to capture the basic ingredients of the relationship between compliance and strain amplification. In a forthcoming publication, we will develop a complete mechanical model based on an auto-coherent description of the heterogeneous glass polymer which goes beyond the approximation of soft/hard separation of the domains and which recovers quantitatively the variation of $Y$ with $J$ while remaining in qualitative agreement with the present mean-field approach.\\
%
Finally, we have shown that it is possible to determine from weakly non linear creep experiments a macroscopic acceleration law (Eq.~\ref{eq:F_1}) with an exponent $n$ close to $2$, in agreement with the the theoretical value predicted by Long~\textit{et al.}~\cite{long_dynamics_2018}. This value is unambiguously determined at the beginning of the creep, when the stress field is not too disordered. However, the local stress disorder increases as creep proceeds which results in an amplification of the stress achieved as suggested by the decrease of $Y$ with $J$.\\
Lastly, our approach allows to derive the following rescaling for compliance in the weakly non-linear creep regime
\begin{equation}
	J_{NL}(\Sigma,t)=J_L \left( t e^{-\left(\frac{A(J)\Sigma}{Y_0}\right)^2} \right),
	\label{eq:Jfin}
\end{equation}
where $A(J)=Y_0/Y(J)$. Using the above expression and the fitted values of $Y(J)$ (green line in Fig.~\ref{subfig:4b}), it was possible to collapse with great accuracy all the non-linear creep data on a single master curve as shown in Fig.~\ref{fig:Jmaster}. In this figure, we have also indicated by arrows the compliance values at which the strain reaches the limiting value used for data processing, \textit{i.e.} 15\%. It emerges that this strain limit corresponds roughly to the reduced time above which the rescaled compliance curves do no longer merge on the master curve, probably as a result of the development of geometrical and material-non linearities which are beyond the scope of this work.\\
%
%
\section{Conclusion}
In this paper we have investigated the weak non-linear mechanical response of an amorphous polymer (PEI) in the vicinity of its glass transition \textit{i.e.} between $T_g-10K$ and $T_g$. This weak non-linear regime is a low stress regime where damage and non-linearities of the strain field are avoided, thus allowing to determine intrinsic material non-linearities. The non linear creep behavior of the polymer in this regime was interpreted as a consequence of the acceleration of the configurational rearrangements of local domains under the action of stress. This acceleration was accounted for by a shift of the relaxation times of the domains. At small compliance values, \textit{i.e.} before the development of substantial stress disorder, the observed exponential dependence of the creep acceleration on applied stress unambiguously demonstrates that the theoretical law predicted by Long~\textit{et al.}~\cite{long_dynamics_2018} describes well the non-linear dependence of the local relaxation time. For larger compliance values, the pre-factor of the stress in the exponential acceleration law was observed to vary. This dependence was attributed to an amplification of the stress applied to the decreasing fraction of domains with slow relaxation times that should carry most of the stress.
From macroscopic creep compliance data, we have been able to estimate the average amplification factor of the local stress versus the macroscopic one. On the basis of a crude mean field approximation which neglects the stress carried by relaxed domains, the linear dependence of the stress amplification factor during the initial stages of creep was qualitatively captured. A more refined mean field description of the weakly non linear response of the polymer would require that the stress-induced shift in the relaxation time distribution is accounted for as creep proceeds, together with local stress distribution. Such a model will be the topic of a forthcoming paper.\\
\section{Dedication}
This article is dedicated to H\'el\`ene Montes, our colleague and friend, who combined great scientific qualities with profound humanity, and who passed away on November $7^{th}$, 2023 while we were working on this project.\\
%
\section{Acknowledgments}
We wish to acknowledge Jean-Claude Mancer for his kind help in the realization of the creep specimens.
%

\begin{figure} [!ht]
	\centering
	\includegraphics[width=0.6\linewidth]{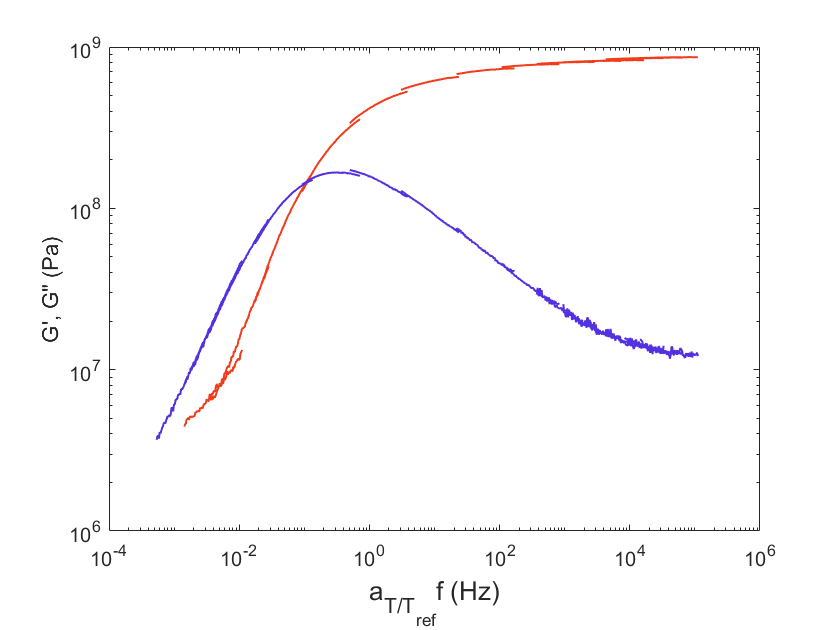}
	\caption{Master curve giving the storage ($G'$, in red) and the loss ($G''$, in blue) shear moduli of the amorphous PEI in the linear range and at the reference temperature $T_{ref}=T_g=213\si{\celsius}$ (shear deformation $\gamma=0.05\%$).}
	\label{fig:linear_master_curve_G}
\end{figure}

\begin{figure} [!ht]
	\centering
	\includegraphics[width=0.6\linewidth]{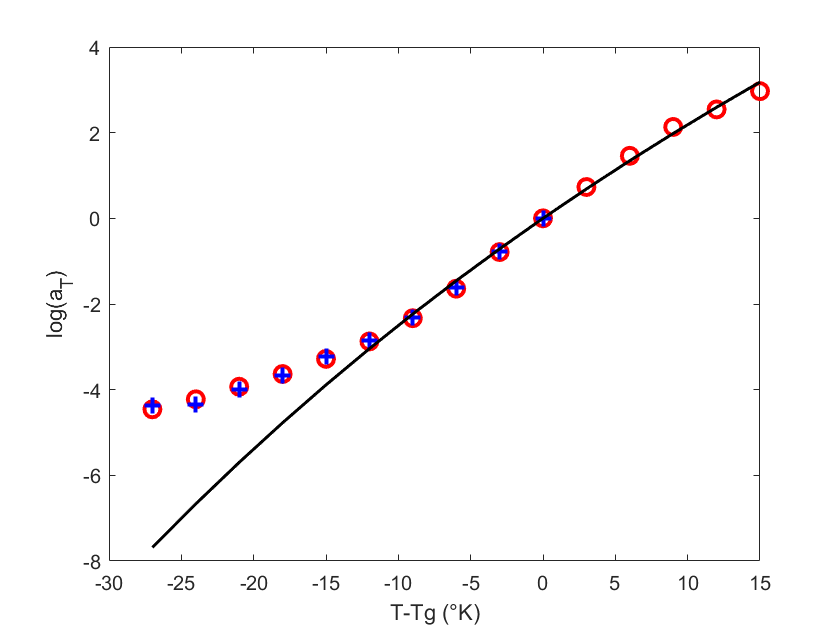}
	\caption{Temperature-dependence of the logarithm of the shift factor $a_T$ determined from time-temperature superposition in the linear regime using either (\textcolor{red}{o}) $G'$ and $G"$ data or (\textcolor{blue}{+}) creep compliance data. $a_T$ values were measured directly from the frequency shifts (resp. the time shifts) of $G'$ and $G"$ (resp. $J(t)$) isotherms. The solid line is a fit of the data to William-Landel-Ferry (WLF) relationship within the temperature range $0\si{\celsius}<T-T_g<15\si{\celsius}$ with $C_1 = 9.56$ and $C_2 = 32.96$~\si{\kelvin}.}
	\label{fig:aT}
\end{figure}

\bibliographystyle{rsc}
\bibliography{weak_non_lin_creep_biblio}
%
\clearpage
	\begin{center}
	\bigskip{Supplementary Information 1, Roman-Faure \textit{et al}}\\
\end{center}
\begin{center}
	\section*{Linear viscoelastic properties of the PEI polymer}
\end{center}
The linear shear viscoelastic properties of the PEI polymer were measured in a torsional mode using rectangular specimens 10x2x2~\si{\milli\meter}$^3$ and a MCR 702 rheometer (Anton Paar, Austria). Fig.~\ref{fig:linear_master_curve_G} shows the resulting master curve obtained at $T_{ref}=213$~$\si{\celsius}$ and for a shear deformation $\gamma=0.05\%$. The corresponding shift factors $a_T$ are reported in Fig.~\ref{fig:aT} as red circles. A fit of the temperature dependence of the shift factors to William-Landel-Ferry (WLF) relationship within the temperature range $0\si{\celsius}<T-T_g<15\si{\celsius}$ provided $C1=9.56$ and $C2=32.96$.\\

In addition, the shift factors determined from the time-temperature superposition of the creep compliance data in the linear range (applied stress $\Sigma=1$~\si{\mega\pascal}) are also reported in Fig.~\ref{fig:aT} as blue crosses for temperatures in between $T_g$ and $T_g-27$~\si{\celsius}. A very good agreement is observed between cyclic shearand tensile creep measurements.\\

%

\clearpage
	\begin{center}
	\bigskip{Supplementary Information 2, Roman-Faure \textit{et al}}\\
\end{center}
\begin{center}
	\section*{Creep experiments: strain recovery sequences}
\end{center}
As detailed in the main text, the building of creep master curves in the linear and non linear regime involved successive creep sequences carried out at the same stress level but increasing temperatures below $T_g$. In order to preserve the accuracy of the strain measurements, successive creep measurements at a given stress level were performed using the same specimen which was maintained fixed within the grips of the rheometer. In between each creep sequence, the specimen was annealed under a zero applied stress at $T-T_g=-3$~\si{\celsius} in order to recover the creep strain. We detail below this strain recovery process both in the linear and in the non linear regime.\\

\begin{figure} [!ht]
	\centering
	\includegraphics[width=0.5\linewidth]{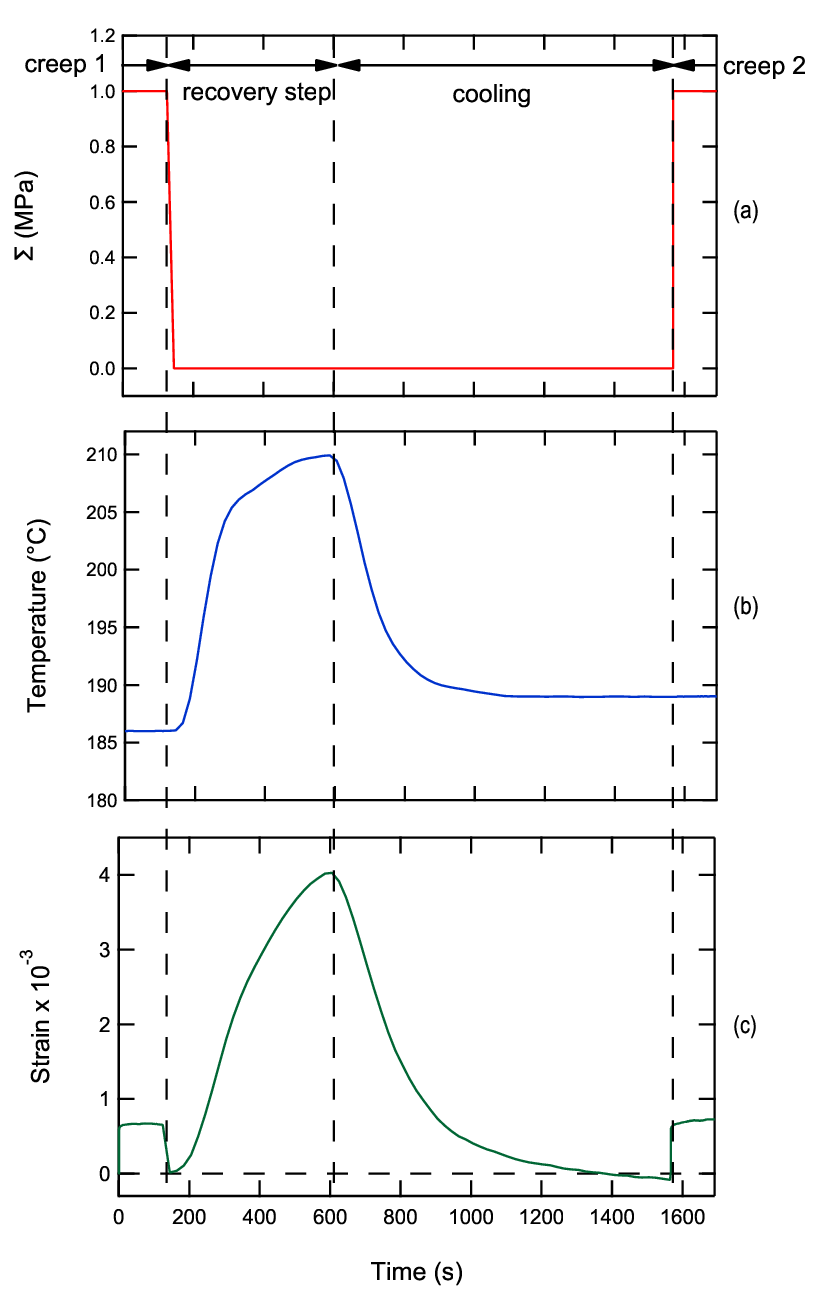}
	\caption{Strain recovery of a PEI specimen after a creep sequence (creep 1) in the linear regime at $T=186$~\si{\celsius} (i.e. $T-T_g=-27$~\si{\celsius}) and under an applied stress $\Sigma=1$~\si{\mega\pascal}. (a) applied stress $\sigma$; (b) temperature; (c) tensile strain. After removal of the applied stress, the specimen is heated up to 210~\si{\celsius} within 460~\si{\second} (recovery step) and subsequently cooled down to the temperature ($T=189$~\si{\celsius}) of the next creep sequence (creep 2).} 
	\label{fig:strain_recov1}
\end{figure}

In Fig.\ref{fig:strain_recov1}, the applied stress, the temperature and the tensile strain are reported as a function of temperature for a creep and recovery sequence in the linear regime ($\Sigma=1$~\si{\mega\pascal}). At the end of the first creep sequence (creep 1), the specimen is heated up to 210~\si{\celsius} and subsequently cooled down to the temperature of the next creep sequence ($T=189$~\si{\celsius}). When considering the change in specimen strain during the recovery step (Fig.~\ref{fig:strain_recov1}c), two opposite effects are at play: one is the thermal expansion of the specimen when it is heated to the recovery temperature, the other one is the contraction of the specimen due to strain recovery. For the considered creep strain, thermal expansion is clearly the predominant effect during the recovery step. However, it can be seen that the specimen has fully recovered its strain after it has been cooled down to the temperature of the next creep step (creep 2 in Fig.\ref{fig:strain_recov1}).\\

\begin{figure} [!ht]
	\centering
	\includegraphics[width=0.5\linewidth]{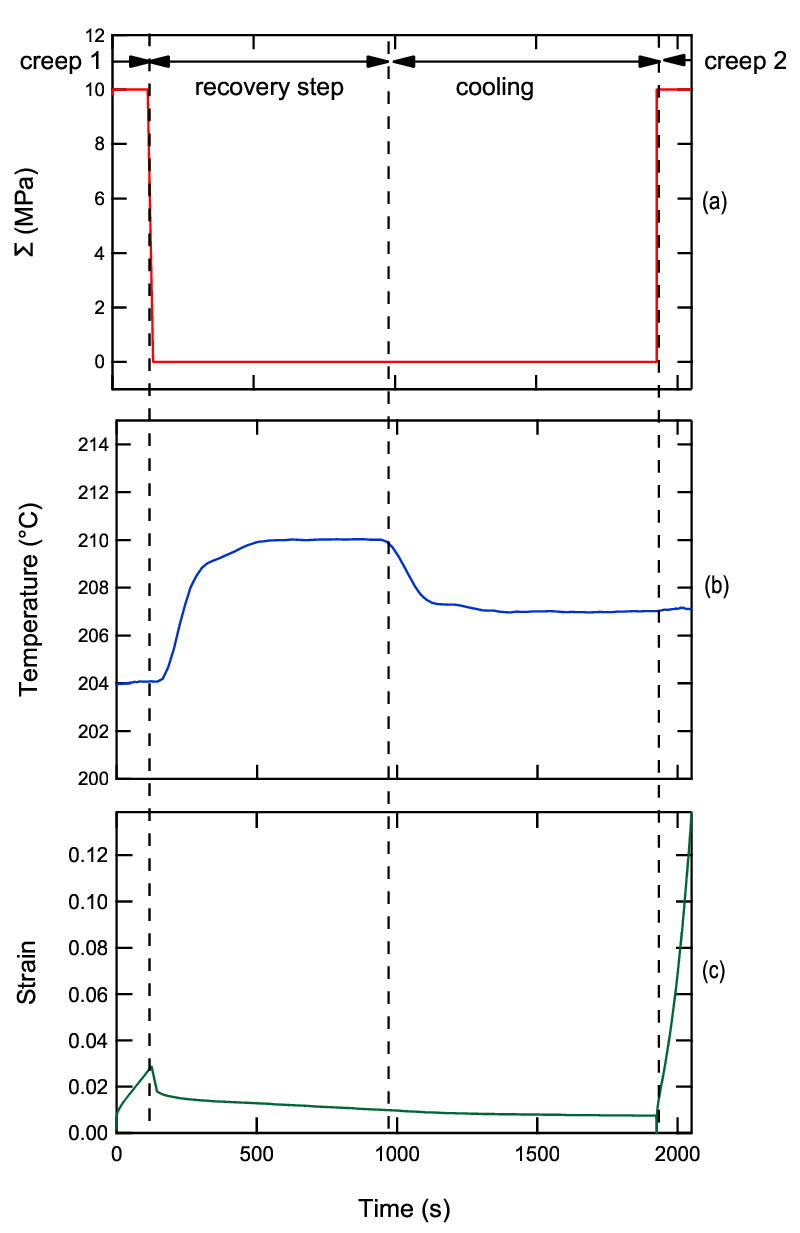}
	\caption{Strain recovery of a PEI specimen after a creep sequence (creep 1) in the non linear regime at $T=204$~\si{\celsius} (i.e. $T-T_g=-9$~\si{\celsius}) and under an applied stress $\Sigma=10$~\si{\mega\pascal}. (a) applied stress $\sigma$; (b) temperature; (c) tensile strain. After removal of the applied stress, the specimen is heated up to 210~\si{\celsius} and allowed to recover during 840~\si{\second} (recovery step) and subsequently cooled down to the temperature ($T=207$~\si{\celsius}) of the next creep sequence (creep 2).} 
	\label{fig:fig:strain_recov2}
\end{figure}

In Fig.\ref{fig:fig:strain_recov2}, similar measurements are reported in the case of creep experiments carried out in the non linear regime ($\Sigma=10$~\si{\mega\pascal}). Here strain recovery predominates over thermal expansion during the recovery state. When the sample is cooled down, a residual strain is remaining but its magnitude (about 0.7\%) is much lower than the creep strain achieved during the subsequent creep sequence ((creep 2 in Fig.\ref{fig:fig:strain_recov2})) at 207~\si{\celsius};\\
\clearpage
	\begin{center}
	\bigskip{Supplementary Information 3, Roman-Faure \textit{et al}}\\
\end{center}
\begin{center}
	\section*{Creep compliance: aging effects}
\end{center}
In this supplementary information, we detail the effects of polymer aging on the creep master curves. Indeed, creep measurements are performed slightly below the glass transition, at temperatures between 186~\si{\celsius} and 213~\si{\celsius}, where aging is significant at the time scale of the experiments. As extensively studied by Struik~[1] and many others, the time dependence of compliance in this aging regime depends on the thermal history of the polymer.\\
In what follows, we assess the influence of the thermo-mechanical history of the specimen during the strain recovery step preceding each creep sequence.For that purpose, master curves were established by increasing the duration of the isotherms preceding the application of creep sequences from 600~\si{\second} (the standard value used for the experiments reported in the main text) to 6000~\si{\second}. In what follows, we will denote as 'young' and 'aged' the specimen subjected to 600~\si{\second} and 6000~\si{\second} isotherms, respectively.\\

\begin{figure} [!ht]
	\centering
	\includegraphics[width=0.6\linewidth]{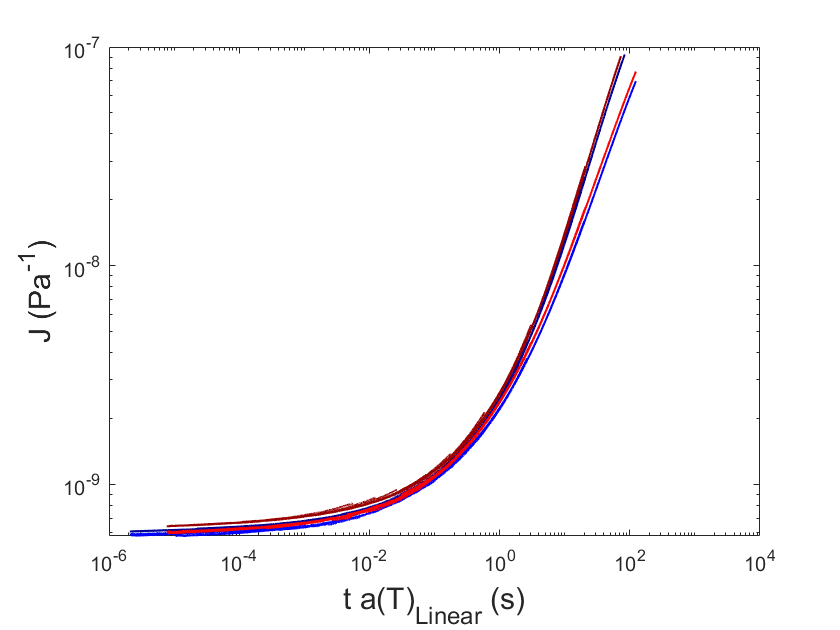}
	\caption{Creep compliance master curves as a function of the thermomechanical history of the specimen for stresses lying in the linear ($\Sigma$=1~\si{\mega\pascal}) and non linear ($\Sigma$=1~\si{\mega\pascal}) ranges. red and purple lines correspond to old and young specimens, respectively, in the linear regime. Light blue and blue curves correspond to old and young specimens, respectively, in the non linear regime.}
	\label{fig:creep_aging1}
\end{figure}

\begin{figure} [!ht]
	\centering
	\includegraphics[width=0.6\linewidth]{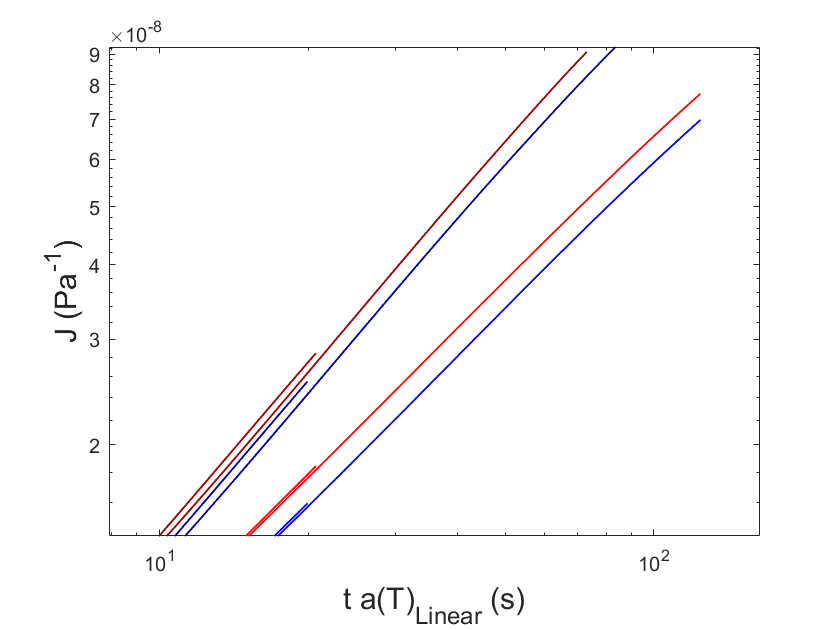}
	\caption{Detail of Fig.~\ref{fig:creep_aging1} for $ta_t \gtrsim  10$~\si{\second}.}
	\label{fig:creep_aging2}
\end{figure}

\begin{figure} [!ht]
	\centering
	\includegraphics[width=0.6\linewidth]{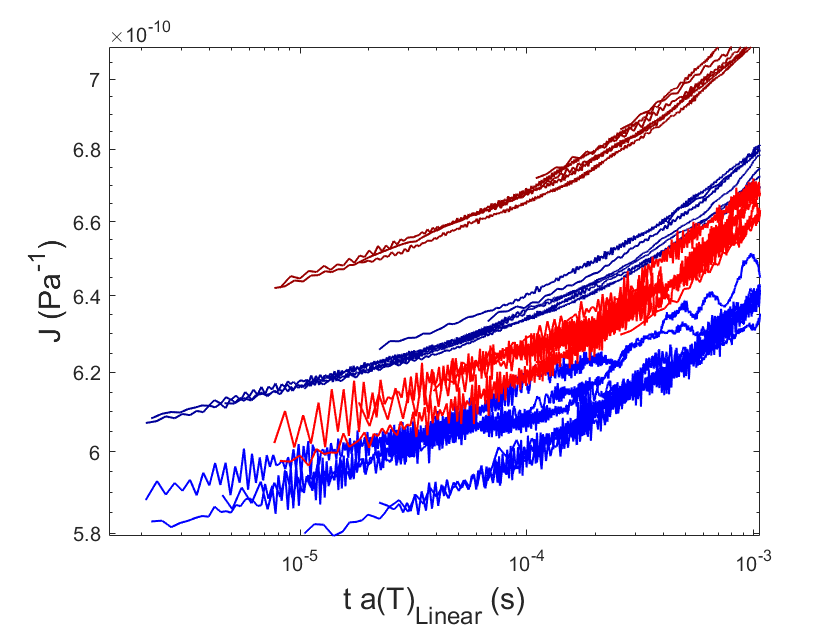}
	\caption{Detail of Fig.~\ref{fig:creep_aging1} for $ta_t \lesssim  10^{-3}$~\si{\second}.}
	\label{fig:creep_aging3}
\end{figure}

In Fig.~\ref{fig:creep_aging1}, creep master curves for young and aged specimens are reported for applied stresses lying in the linear ($\Sigma=$1~\si{\mega\pascal}) and in the non linear range ($\Sigma=$5~\si{\mega\pascal}). Details of these creep compliance curves are provided in Figs.~\ref{fig:creep_aging2} and \ref{fig:creep_aging3} for long times (\textit{i.e.} high temperatures) and small times (\textit{i.e.} low temperatures), respectively.\\

In the long time regime ($ta_t \gtrsim  10$~\si{\second}, Fig~\ref{fig:creep_aging2}), a shift of the master curve of the young polymer to low time is observed  as compared to the old one, as expected from the work by Struik~[1]. However, the magnitude of this shift is lower than that resulting from the application of a stress in the non linear regime. In addition, it occurs to roughly the same extent at all $J(t)$ values. As a consequence, its effect on the determination of the acceleration factor $F$ can be neglected.

Conversely, at low times ($ta_t \lesssim  10^{-2}$~s) aging has a much more pronounced effect on the creep compliance response. As shown in Fig.~\ref{fig:creep_aging3}, the shift of the compliance curves as a result of aging is of the same order of magnitude than that resulting from the application of stresses in the non linear range. Although interesting, a detailed description of the acceleration factor $F$ in this aging regime is beyond the scope of this study. As a consequence, all experimental creep data for $J<3 \;10^{-9}$~\si{\mega\per\pascal} were discarded for the determination of the acceleration factor $F$.\\

\subsection*{References}

Struik, L.C.E., \textit{Physical Aging in Amorphous Polymers and Other Materials}, Elsevier, 1980

\begin{figure} [!ht]
	\centering
	\includegraphics[width=0.5\linewidth]{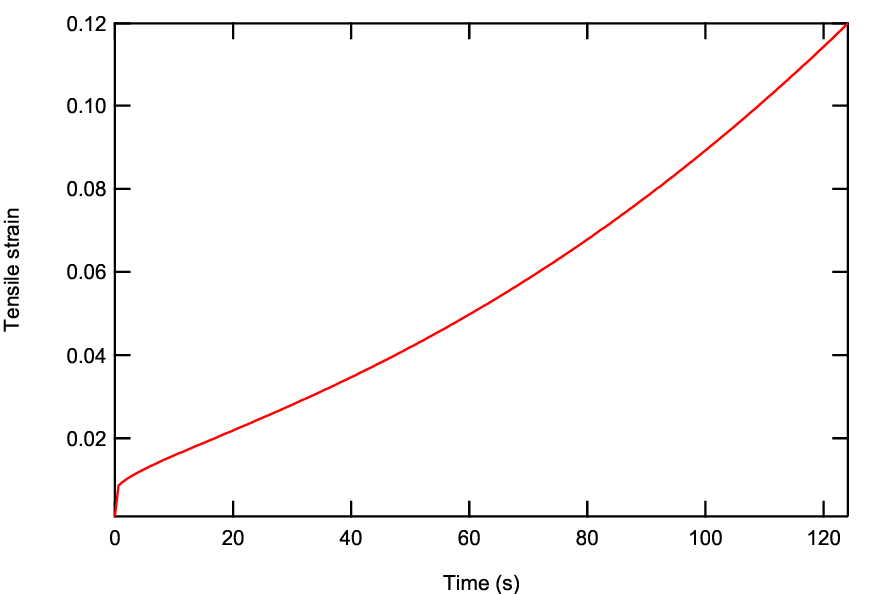}
	\caption{Tensile strain as a function of time during creep at 207~\si{\celsius} under an applied stress of 10~\si{\mega\pascal}.}
	\label{fig:strain}
\end{figure}

\begin{figure}  [!ht]
	\centering
	\includegraphics[width=0.4\linewidth]{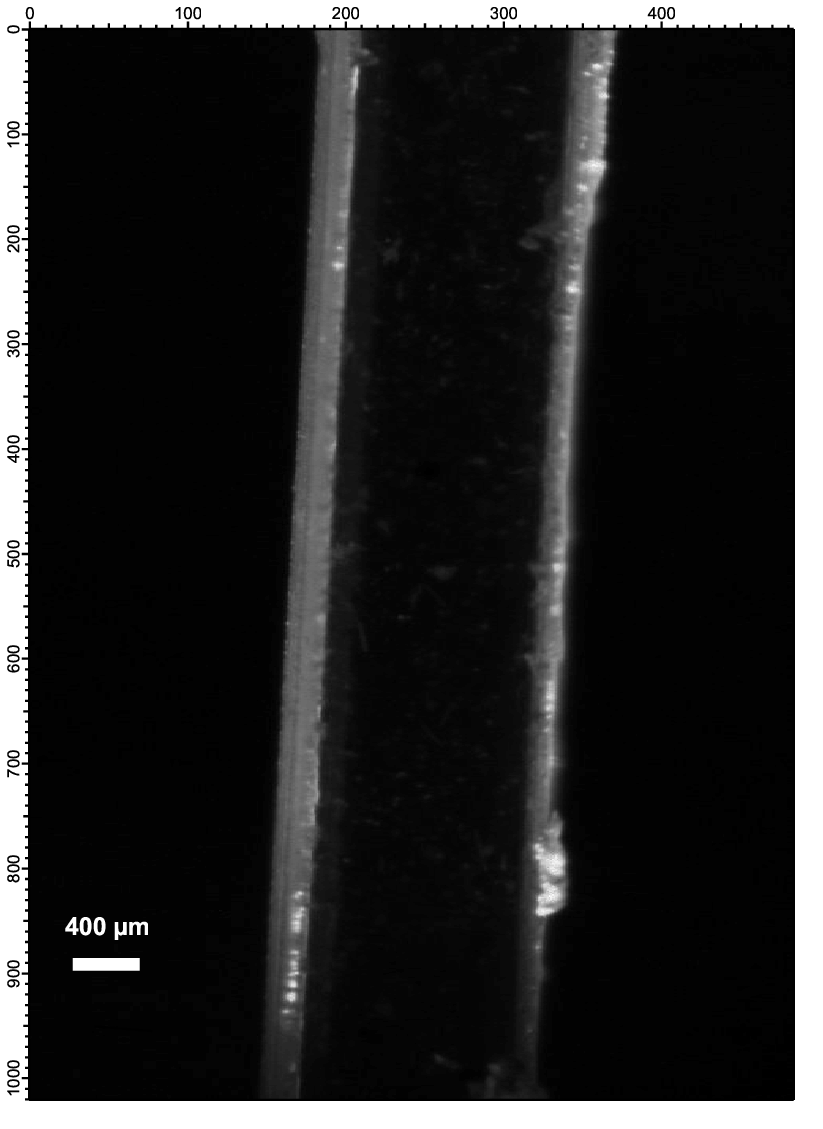}
	\caption{Video recording of a PEI specimen during a tensile test at 207~\si{\celsius} under an applied stress of 10~\si{\mega\pascal}}.
	\label{fig:video}
\end{figure}

\clearpage
	\begin{center}
	\bigskip{Supplementary Information 4, Roman-Faure \textit{et al}}\\
\end{center}
\begin{center}
	\section*{In situ video recording of a PEI specimen during a creep experiments}
\end{center}
The video depicts a video recording of a PEI specimen during a creep test carried out at 207~\si{\celsius} under an applied stress of 10~\si{\mega\pascal}. Video were acquired through the windows of the rheometer oven using a zoom lens and a CCD camera. The corresponding time-dependent strain is shown in Fig.~\ref{fig:strain}. The frame rate for image acquisition 1.6~\si{\hertz} and the image resolution is 7.6~\si{\micro\meter}/pixel.\\
No necking nor crazing were evidenced within the limits of optical resolution (see Fig.\ref{fig:video}).\\
\end{document}